\documentclass[aps,pra,twocolumn,groupedaddress,nofootinbib, longbibliography]{revtex4-2}

\usepackage{color}
\usepackage{dcolumn}
\usepackage{graphicx}

\usepackage{amsmath,amsfonts,amssymb,amsthm}

\usepackage{bbold}
\usepackage{dsfont}

\usepackage{soul}

\usepackage{fancyhdr}
\usepackage{multirow}
\usepackage{bm}
\usepackage{dsfont}
\usepackage{cancel}

\usepackage{xifthen}

\usepackage{float}
\graphicspath{{"./graphs/"}{"./scheme/"}}

\usepackage{graphicx}
\usepackage[usenames,dvipsnames,dvipsnames,svgnames]{xcolor}

\usepackage[colorlinks,bookmarks=false,
linkcolor=blue,
urlcolor=blue,
citecolor=blue
]{hyperref}

\usepackage{xspace}

\def\mean#1{\ensuremath{\langle #1 \rangle}}

\def\dissip#1{\mathcal{D}\big[#1\big]}

\def\commut#1#2{\lbrack #1,#2 \rbrack}

\renewcommand{\H}{\mathcal{H}}

\def\op#1{\hat{#1}}

\renewcommand{\ao}[1][]{%
	\ifthenelse{\equal{#1}{}}{\ensuremath{\op{a}}}{\ensuremath{\op{#1}}}%
}
\newcommand{\co}[1][]{%
	\ifthenelse{\equal{#1}{}}{\ensuremath{\op{a}^{\dagger}}}{\ensuremath{\op{#1}^{\dagger}}}%
}

\usepackage{array,calc}

\usepackage[english]{babel}
\usepackage{pgfplots} 
\usetikzlibrary{plotmarks}
\newcommand{\me}{\mathrm{e}}
\newcommand{\GP}{Gross-Pitaevskii}

\newcommand{\TW}{truncated Wigner}
\newcommand{\Ntraj}{\ensuremath{N_{\text{traj}}}\xspace}

\usepackage{scalerel}
\newcommand\Tau{\mathcal{T}}

\usepackage[normalem]{ulem}

\begin{document}

\title{Quantum dynamics of Dissipative Kerr solitons}

\author{Kilian Seibold}
\affiliation{Institute of Physics, Ecole Polytechnique F\'ed\'erale de Lausanne (EPFL), CH-1015, Lausanne, Switzerland}
\email{kilian.seibold@epfl.ch}
\author{Riccardo Rota}
\affiliation{Institute of Physics, Ecole Polytechnique F\'ed\'erale de Lausanne (EPFL), CH-1015, Lausanne, Switzerland}
\author{Fabrizio Minganti}
\affiliation{Institute of Physics, Ecole Polytechnique F\'ed\'erale de Lausanne (EPFL), CH-1015, Lausanne, Switzerland}
\author{Vincenzo Savona}
\affiliation{Institute of Physics, Ecole Polytechnique F\'ed\'erale de Lausanne (EPFL), CH-1015, Lausanne, Switzerland}

\date{\today}

\begin{abstract}
Dissipative Kerr solitons arising from parametric gain in ring microresonators are usually described within a classical mean-field framework. Here, we develop a quantum-mechanical model of dissipative Kerr solitons in terms of the truncated Wigner method, which accounts for quantum effects to lowest order. We show that the soliton experiences a finite lifetime due to quantum fluctuations originating from losses. Reading the results in terms of the theory of open quantum systems, allows to estimate the Liouvillian spectrum of the system. It is characterized by a set of eigenvalues with finite imaginary part and vanishing real part in the limit of vanishing quantum fluctuations. This feature shows that dissipative Kerr solitons are a specific class of dissipative time crystals. 
\end{abstract}
	
\maketitle
	
\section{Introduction}
\label{sec:intro}

Kerr frequency combs (KFCs) \cite{Chembo2016, Kues2019, Karpov2019, Kovach20, Scott2020, Kippenbergeaan8083} are optical frequency combs generated by driving high-Q Kerr–nonlinear optical microresonators with a single-frequency continuous-wave laser \cite{DelHaye2007, Kippenberg555}. 
By driving sufficiently above a power threshold determined by the Kerr nonlinearity, and under appropriate conditions for the dispersion of the microresonator optical modes, the parametric process generates a comb of evenly spaced peaks in the frequency spectrum \cite{Udem2002,Gaeta2019}. 
Since the first demonstration of KFCs \cite{DelHaye2007}, they have been observed countless times in a variety of platforms, materials, and spectral ranges, including silica microtoroid resonators \cite{Yi:15, Ma2019}, crystalline microresonators \cite{PhysRevLett.101.093902}, silicon nitride waveguide resonators \cite{Levy2010, Ferdous2011, Okawachi:11, Joshi:16, Brasch2016}, diamond \cite{Hausmann2014}, aluminum nitride \cite{Jung:13, Jung:14}, lithium niobate \cite{He:19, Wang2019}, and silicon \cite{Griffith2015}. 

KFCs emerge from multiple parametric resonant four-wave mixing processes.
On the one hand, they result from a double balance process, where the nonlinear frequency shifts are balanced by the mode-frequency dispersion in the microresonator. On the other, the cavity losses are balanced by the gain induced by the continuous-wave driving field. 

For sufficiently strong drive, the frequency spacing in the comb can be as small as the free spectral range of the microresonator. In this case, a bright pulse circulating within the resonator, called dissipative Kerr soliton (DKS) is formed \cite{Leo2010, Herr2014, Chembo2016, Weiner2017, ThesisEPFLKarpov, Gaeta2019,Lobanov20}. DKSs are time-periodic solutions of an otherwise time-independent open quantum system dynamics \cite{GreluPhilippe2016Nocd, LugiatoLuigi2015Nos, Kippenbergeaan8083}. 
A notable feature of DKSs is that they are dynamically stable: their waveform retains its shape \textit{indefinitely}, making DKSs a promising resource for precision measurements \cite{Kippenbergeaan8083, Kippenberg555}, time keeping \cite{Papp:14, Gerginov:05}, frequency metrology \cite{PhysRevX.3.031003, PhysRevLett.101.053903, PhysRevLett.109.263901, Spencer2018}, pulse shaping \cite{Ferdous2011}, communication engineering \cite{Pfeifle2014, PhysRevLett.114.093902, Marin-Palomo2017}, high-resolution spectroscopy \cite{Bernhardt2010, yu2018, PhysRevLett.96.153001, PhysRevA.80.053806, PhysRevLett.111.023007, Morgenweg2014,Cingoz2012, Suh2016}, and quantum information processing \cite{Reimer2016}. 

The rapid development of miniaturized integrated systems for KFCs and DKSs, operating at low power where quantum effects are expected to be relevant, calls for a detailed study of the influence of quantum fluctuations on the spectral and dynamical properties of DKSs in the low-power regime. While the quantum properties of KFCs have been extensively investigated \cite{PhysRevLett.107.030505, Roslund2014, PhysRevLett.112.120505, Reimer2016, Kues2017, Imany:18, Kues2019, Yang2021}, only recently the quantum mechanical properties of the DKS regime have been experimentally addressed \cite{guidry2021quantum}. In addition, both in the case of KFCs operated below the parametric oscillation threshold and for DKSs, quantum effects have been modeled under the assumption of linearized quantum fluctuations, resulting in Gaussian quantum fields \cite{PhysRevA.93.033820, PhysRevLett.107.030505, Roslund2014, PhysRevLett.112.120505, Reimer2016, Kues2017, Imany:18, Kues2019, Yang2021, guidry2021quantum}.

Here, we investigate the dynamics of DKSs using the truncated Wigner approximation \cite{GardinerZollerQNoise, walls2008quantum, Vogel89, Opanchuk13, PhysRevA.96.043616, Polkovnikov2010} -- an approximation to model driven-dissipative quantum systems based on the Wigner quasi-probability distribution that reliably describes small quantum fluctuations.
Our results show that for any finite value of the input power in the microresonator, the DKS solution does not persist indefinitely but decays over time due to the presence of nonlinearity and fluctuations induced by losses. At long times, a nonequilibrium steady state that restores the time-invariant symmetry of the system eventually emerges. We demonstrate that the timescale for this steady state to be reached has a power-law behavior depending on the strength of the nonlinearity, and a true DKS emerges only in the limit of vanishing nonlinearity and infinite driving field amplitude.

The study of the DKS dynamics in terms of the theory of open quantum systems shows that DKSs are a specific manifestation of a dissipative time crystal (DTC\footnote{In this work, DTC stands for dissipative time crystal and not for discrete time crystal \cite{Zhang2017}. In the context of open quantum systems, DTCs are also called boundary time crystals \cite{Iemini2018}.}).
DTCs \cite{Sacha_2017,  Nakatsugawa2017, Wang2018, Tucker2018, Gong2018, Owen_2018, Iemini2018, Lledo2019, osullivan2019dissipative,  Gambetta2019, Nalitov2019, Kessler2019, Zhu_2019, Lledo2019, Dogra2019, Chiacchio2019, Buca2019PRL, Buca2019Nature, Lled2020, Laz2020, Chinzei2020, Kessler_2020} are a peculiar phase of a driven-dissipative quantum system where the time-translational symmetry of the equation of motion is broken and non-stationary long-lived states spontaneously occur \cite{Booker_2020}. 
In the past years, intense research has been devoted to investigating the conditions under which dissipation can prevent a quantum many-body system from reaching a stationary state \cite{Buca2019Nature, Buca2019PRL, Booker_2020,Engelhardt2021}. This has led to numerous proposals of quantum systems supporting a DTC phase \cite{Iemini2018,Tucker2018,Barbarena2019,Cosme2019,Lledo2019,Tindall19,Seibold2020,Buca2019PRL,HurtadoGutierrez2020,Prazeres2021,minganti2020correspondence, Kess2021}. 
DTCs admit a natural explanation in terms of the eigenvalues of the Liouvillian super-operator, which generates the time evolution of the density matrix of an open quantum system \cite{Iemini2018,Lidar2020lecture}. In a DTC, multiple eigenvalues of the Liouvillian exist with vanishing real and finite imaginary part \cite{Albert14, PhysRevX.6.041031, Baumgartner_2008}, giving rise to a nonstationary dynamics with diverging relaxation time towards a steady state. 
Here, we study the properties of the Liouvillian spectrum of a Kerr microresonator, showing that the long-lived DKS is associated with a DTC. More specifically, the Liouvillian is characterized by a set of eigenvalues whose imaginary parts are integer multiples of the frequency defining the free spectral range of the microresonator, and whose real part goes to zero in the thermodynamic limit of infinite photon number and vanishing nonlinearity.

The paper is organized as follows. In Sec.~\ref{sec:theoretical framework}, we survey the theoretical framework used for the quantum analysis of DKSs. In Sec.~\ref{sec:results}, we discuss the result obtained for the dynamics of the system: in particular, we compute the Liouvillian gap for decreasing drive power and depict a schematic representation of the spectrum of the Liouvillian. The main findings and conclusions of this work are drawn in Sec.~\ref{sec:conclusions}.

\section{Theoretical framework}
\label{sec:theoretical framework}
	
\subsection{The open quantum system model and Liouvillian gap}
\label{subsec:ME and Liouvillian gap}

\begin{figure}
	\centering
	\includegraphics[width=1\linewidth]{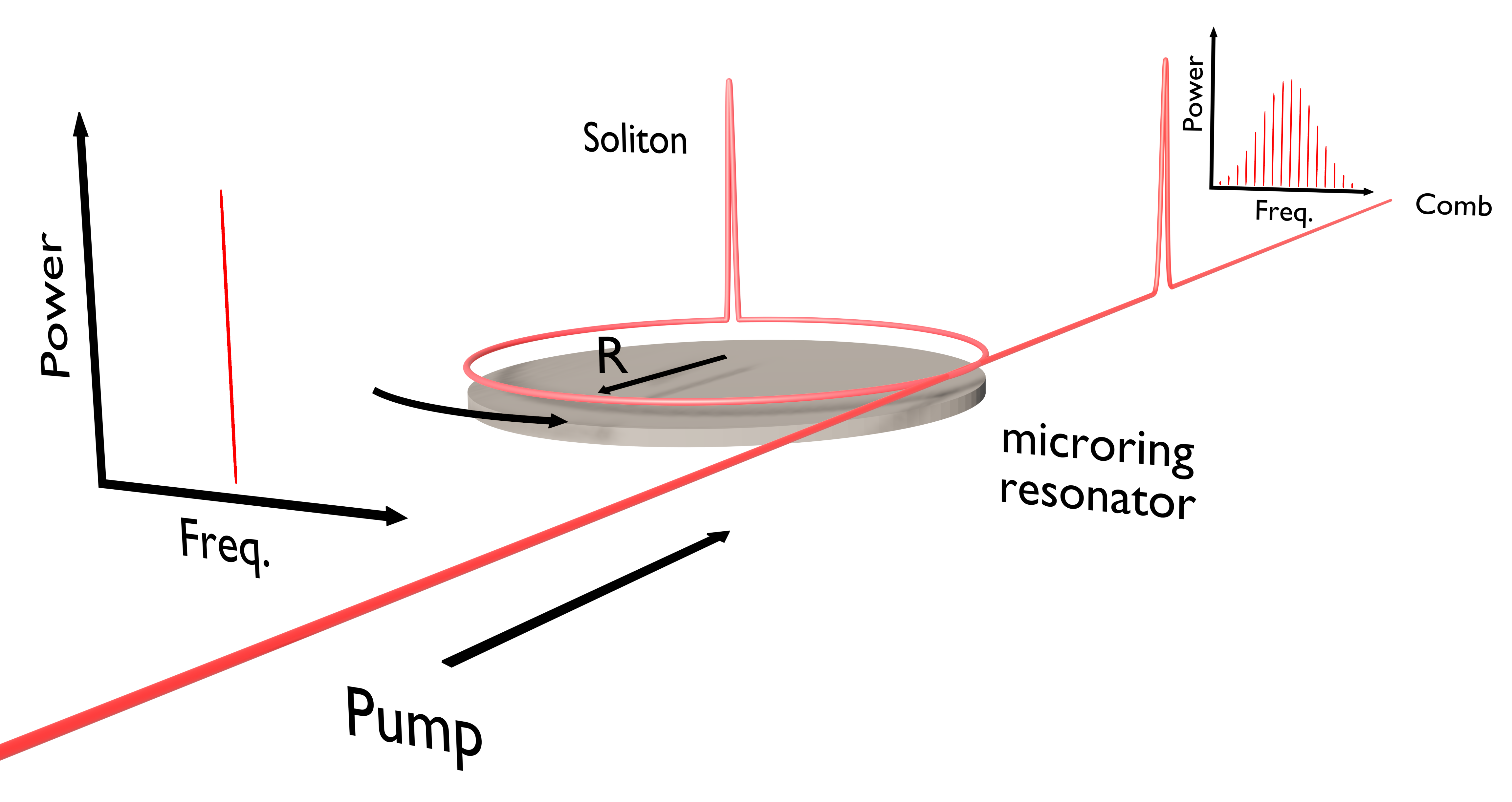}
	\caption{Schematic representation of the generation of a Kerr optical frequency comb using a high-Q Kerr optical ring microresonator. A continuous wave source drives the ring, which induces the propagation of a soliton (depicted in red) along the ring. The output signal shows the optical frequency comb. When all the resonator modes participate in the parametric process, the non-linear dynamics give rise to a DKS.}
	\label{fig:schema ring}
\end{figure}

We consider a driven high-Q continuous optical ring microresonator, whose schematic is shown in Fig.~\ref{fig:schema ring}.
The system Hamiltonian, in a frame rotating at the driving frequency, reads ($\hbar=1$):
\begin{equation}
	\begin{split}
		\op{\H} =& \sum_l\sigma_l\co_l\ao_l + \dfrac{\kappa}{2}F(\co_0+\ao_0)\\ &+ \dfrac{g}{2}\sum_{m,n,p,q}\delta_{m+p,n+q}\co_n\co_q\ao_m\ao_p\,,
	\end{split}
	\label{eq: Hamiltonian ring resonator}
\end{equation}
where $\ao_l$ ($\co_l$) is the annihilation (creation) operator of the $l$-th angular momentum mode (i.e., the discrete set of whispering gallery modes), satisfying the commutation relation $\left[\ao_j,\co_k\right]=\delta_{jk}$. 
Only the lowest-energy mode of the microresonator ($l=0$) is driven by an external continuous-wave laser of amplitude $F$. 
The Kerr interaction strength can be obtained from a microscopic model as $g = \hbar\omega_0^2c n_2/(n_0^2 A_{\rm eff}L)$ \cite{PainterOE2005, PhysRevApplied.12.064065, PhysRevA.93.033820}, where $c$ is the speed of light in vacuum, $n_0$ is the refractive index of the medium at the fundamental resonator frequency $\omega_0$, $n_2$ is the Kerr parameter, $A_{\rm eff}$ is the effective mode area, and $L$ is the resonator length.
Note that miniaturizing the resonator means decreasing the effective mode volume ($V_{\rm eff}=A_{\rm eff} L$) and, hence, increasing the Kerr interaction strength $g$.
We set $\sigma_l = \sigma_0+\omega_0-\omega_l$, where $\sigma_0 = \omega_p - \omega_0$ is the detuning between the driving frequency $\omega_p$ and the fundamental resonator frequency $\omega_0$, and $\omega_l$ is the dispersion relation (which in this work is assumed parabolic, $\omega_l\propto l^2$, see also Fig.~\ref{fig:integrated disp relation}).

In order to account for the finite lifetime of the photons inside the microresonator, we describe the dynamics of the open system in terms of its reduced density matrix $\op{\rho}$. 
Assuming a weakly interacting and a memoryless environment (i.e., Born and Markov approximations), $\op{\rho}$ solves the Lindblad quantum master equation \cite{OpenQBreuer,GardinerZollerQNoise}:
\begin{equation}
	\frac{d\op{\rho}}{dt}=\mathcal{L}\op{\rho} = -i\commut{\op{\H}}{\op{\rho}}+ \kappa \sum_{l}\dissip{\ao_l}{\op{\rho}}\,.
	\label{eq:master equation}
\end{equation}
Here $\mathcal{D}\big[\ao_l\big]\op{\rho} = \ao_l\op{\rho}\ao_l^{\dagger}-1/2(\ao_l^{\dagger}\ao_l\op{\rho}+ \op{\rho}\ao_l^{\dagger}\ao_l)$ is the dissipator in Lindblad form accounting for the loss of photons from a mode $l$ into the environment and $\kappa$ is the dissipation rate (which we assume uniform).
$\cal L$ is the Liouvillian superoperator and its spectrum, defined by the equation $\mathcal L \hat \rho_j =  \lambda_j \hat \rho_j$, encodes the full dynamics of an open quantum system. 
In most physically relevant cases, the Liouvillian superoperator $\cal L$ has a unique zero eigenvalue, which defines the nonequilibrium steady state $d\hat{\rho}_{ss}/dt=0$ \cite{Nigro_2019, Minganti2018}. 
All other eigenvalues of the Liouvillian have a negative real part, determining the irreversible dissipative dynamics towards the steady state. 
The eigenvalue $\lambda_j$, whose real part is the smallest nonzero in modulus, defines the Liouvillian gap $\Lambda$. 
$\Lambda$ corresponds to the inverse of the longest relaxation timescale of the system. Critical phenomena, such as dissipative phase transitions or the emergence of DTCs, are associated with a closure of the Liouvillian gap, i.e., $\Lambda \to 0$.
A  complete account of the spectral theory of Liouvillians can be found in, e.g., Ref. \cite{Minganti2018}.

\subsection{Classical-field approach to solitons}

To numerically simulate the optical ring microresonator, we will consider a finite number of modes $N_m$ around the $l=0$ driven mode.
Except if differently specified, we will set $N_m=101$ (i.e., we consider only the modes $l=[-50,-49,...,50]$). We verified that the results shown hereafter are only affected in one part in $10^{5}$ on the total population, as a result of this truncation in the number of modes. 
Despite this simplification, the numerically exact solution of the master equation \eqref{eq:master equation}, in the regime of large occupation considered here, would be computationally unfeasible.

	DKSs in the weakly nonlinear regime are usually modeled in terms of the classical \GP{} (GP) equation, whereby the classical field amplitudes of the resonator modes $\alpha_l = \langle\hat{a}_l\rangle$ obey the equation
	\begin{equation}
	\begin{split}
		\alpha_{l}(t+dt) & =  \alpha_{l}(t) + i \left\{\left( \sigma_l + i\frac{\kappa}{2}\right) \alpha_{l}(t) \right.\\
		&  +g\sum_{m,n,p}\delta_{n+l,m+p} \alpha_{m}(t) \bar{\alpha}_{n}(t) \alpha_{p}(t) \\
		&  \left. -\frac{\kappa}{2} F \delta_{l,0} \right\} dt\;,
	\end{split}
	\label{eq: GPE}
\end{equation}	
where $\bar{\alpha}_{n}$ indicates the complex conjugate of the field $\alpha_{n}$. The GP equation leads directly to the \textit{Lugiato-Lefevre equation} \cite{PhysRevLett.58.2209, PhysRevA.89.063814} describing the real-space dynamics of the soliton.

Note that Eq.~\eqref{eq: GPE} is invariant under the scaling relation 
\begin{equation}
	\tilde{\alpha}_{l} = \alpha_{l}/\sqrt{\tilde{N}}, \quad
	\tilde{g} = g\tilde{N}, \quad
	\tilde{F} = F /\sqrt{\tilde{N}},
\label{eq: rescaling relation}
\end{equation}
where we introduced the dimensionless scaling parameter $\tilde{N}$.
The GP solution for the rescaled field $\tilde{\alpha}_{l}$ only depends on the product $\tilde{F}^2 \tilde{g}=F^2 g$.
In what follows, all reasults are obtained by setting $\tilde{F}^2 \tilde{g}=1$, which corresponds to a case well above the threshold for soliton formation (see discussion below).

\subsection{The truncated Wigner approximation}

Theoretical studies of DKSs beyond the GP approximation have been mostly carried out by assuming linearized quantum fluctuations around the GP solution, i.e., Gaussian quantum fields \cite{PhysRevA.93.033820, PhysRevLett.107.030505, Roslund2014, PhysRevLett.112.120505, Reimer2016, Kues2017, Imany:18, Kues2019, Yang2021, guidry2021quantum}.
Quantum mechanical properties of the ring resonator are better described with methods based on quasi-probability distributions \cite{GardinerZollerQNoise, walls2008quantum}, such as the \TW{} approximation (TWA).
Indeed, in cases where the quantum effects are a small (but non-negligible) correction to the classical limit of very large photon occupation, these methods account for non-Gaussian quantum fluctuations, which become relevant when increasing $g$.
Below, we recall the main ideas behind the TWA; for a more detailed derivation, we refer the interested reader to Refs.~\cite{GardinerZollerQNoise, walls2008quantum, Vogel89, Opanchuk13, PhysRevA.96.043616, Polkovnikov2010}.

For a single mode of the electromagnetic field, the Wigner quasi-probability distribution function $W(\alpha)$ of a given quantum state expresses the quasi-probability distribution function in the phase space spanned by $Q$ and $P$, with $\alpha=(Q+iP)/\sqrt{2}$.
The quantities $Q$ and $P$ are the (real) eigenvalues of the electromagnetic field quadratures $\hat{q}$, and $\hat{p}$, with $\hat{a}=(\hat{q}+i\hat{p})/\sqrt{2}$. For $N_m$ modes, the Wigner function $W(\vec{\alpha})$ is easily generalized in terms of $N_m$ complex fields $\vec{\alpha}=\{\alpha_{-N_m/2}, \dots \alpha_{N_m/2} \}$. The density matrix can be expressed in terms of the Wigner function as \cite{Glauber69}
\begin{equation}
     W(\vec{\alpha})  = \left(\frac{2}{\pi}\right)^{N_m} \operatorname{Tr}\left[ \prod_{l=-N_m/2}^{N_m/2} \hat{D}(\alpha_{l}) e^{i \pi \hat{a}_l^{\dagger} \hat{a}_l}  \hat{D}(-\alpha_{l})\hat{\rho}\right],
\end{equation}
where $\hat{D}(\alpha_{l})= \exp(\alpha_l \hat{a}_l^\dagger - \alpha_l^* \hat{a}_l)$ is the displacement operator.
$W(\vec{\alpha})$ is a quasi-probability because it is real-valued, but it can take negative values.
The Lindblad master equation for the density matrix of a quantum optical system characterized by a Kerr nonlinearity maps onto a third-order differential equation for $W(\vec{\alpha})$ in the variables $\vec{\alpha}$. The exact solution of this equation is as cumbersome as the solution of the corresponding master equation. However, when in presence of a small Kerr nonlinearity $g$ and for sufficiently well-behaved functions, the third-order terms can be neglected, resulting in the TWA \cite{Vogel89, PhysRevA.96.043616}. The TWA correctly describes quantum fluctuations up to the lowest (i.e. second) order in $\hbar$ with respect to the mean-field equation, holding in the limit of very large photon occupation \cite{Polkovnikov2010}. 

The advantage of the TWA is that it defines a Fokker-Plank equation for the complex fields $\vec{\alpha}$. By choosing an appropriate initial distribution, the Fokker-Plank equation for $W(\vec{\alpha})$ can be cast into a set of Langevin equations for the corresponding stochastic processes on the complex fields $\alpha_{l, \mu}(t)$.
In the case of the DKS model considered here, the Langevin equations read
\begin{equation}
	\begin{split}
		\alpha_{l, \mu}(t+dt) & =  \alpha_{l, \mu}(t) + i \left\{\left( \sigma_l + i\frac{\kappa}{2} - g\right) \alpha_{l, \mu}(t) \right. \\
		& +g\sum_{m,n,p}\delta_{n+l,m+p} \alpha_{m, \mu}(t) \bar{\alpha}_{n, \mu}(t) \alpha_{p, \mu}(t) \\
		& \left. -\frac{\kappa}{2} F\delta_{l,0} \right\} dt + \sqrt{\kappa dt/2}\chi_{l, \mu}(t) \ .
	\end{split}
	\label{eq:TWA_Riccardo}
\end{equation}
Here, the index $\mu$ runs on the distinct Langevin trajectories, and the term $\chi_{l, \mu}(t)$ is a complex Gaussian stochastic variable defining each specific trajectory, and characterized by correlation functions $\mean{\chi_l(t)\chi_l(t')}=0$ and  $\mean{\chi_l(t)\bar{\chi}_{l'}(t')}=dt \,\delta_{l,l'}\delta(t-t')$.
The noise terms $\chi_{l, \mu}(t)$ therefore account for the quantum fluctuations induced by photon losses.

Within the TWA, it is possible to obtain the expectation value of symmetrized product of operators in terms of an average over the sampled Langevin trajectories, according to the formula 
\begin{equation}
	\textrm{Tr} \left[  \op{\rho}(t) \left\{(\co_j)^n,(\ao_k)^m\right\}_{\text{sym}}\right]
	=\langle
	\left(\bar{\alpha}_{j}(t)\right)^n
	\left(\alpha_{k}(t)\right)^m\rangle_{\text{stoch}}\,,
\end{equation}
where we use the notation for the stochastic average $\langle\alpha_{j}(t)\rangle_{\text{stoch}}=
\left(\sum_{\mu=1}^{\Ntraj}\alpha_{j,\mu}(t)\right)/\Ntraj$.
In other words, the expectation value of any observable is obtained by sampling a sufficiently large number \Ntraj of Langevin trajectories, thus recovering the results of the Fokker-Plank equation associated to the TWA. In the following, the convergence of the results with respect to the number of considered trajectories \Ntraj used for the averaging is carefully checked.

From the solutions of the Langevin equations, the number of photons in each mode $l$ of the microresonator is expressed by
\begin{equation}
	N_l(t) = \langle|\alpha_{l}(t)|^2\rangle_{\text{stoch}}-\frac{1}{2}	\;,
	\label{eq:TWAPhotonsMode_l}
\end{equation}
and the photon density at position $\theta$ is obtained as
\begin{equation}
	N_\theta(t) = \langle|\psi_\mu(\theta,t)|^2\rangle_{\text{stoch}}-\frac{N_m}{4 \pi}	\;,
	\label{eq:TWAPhotonsrealspace}
\end{equation}
where 
\begin{equation}
	\psi_\mu(\theta,t) =\dfrac{1}{\sqrt{2 \pi}}\sum_l e^{i\theta l}\alpha_{l,\mu}(t)\,.
	\label{eq:continous spacial field}
\end{equation}
Notice that Eqs.~\eqref{eq:TWAPhotonsMode_l} and \eqref{eq:TWAPhotonsrealspace} clearly illustrate how quantum fluctuations are approximately described by the TWA. In particular, Eq. \eqref{eq:TWAPhotonsMode_l} shows that the classical field modeled by the Langevin equation contains quantum fluctuations corresponding to half a photon per mode. 
Similarly, Eq. \eqref{eq:TWAPhotonsrealspace} suggests that the Langevin field a discrete element of real space $\Delta\theta$, defined by the momentum cutoff introduced by truncating to $m$ modes, contains quantum fluctuations corresponding to $N_m/4\pi$ photons.

Contrarily to the GP equation, Eq.~\eqref{eq:TWA_Riccardo} is not invariant under the rescaling introduced in Eq.~\eqref{eq: rescaling relation}.
Indeed, applying the same rescaling to Eq.~\eqref{eq:TWA_Riccardo}, one obtains
	\begin{equation}
		\begin{split}
			\tilde{\alpha}_{l, \mu}(t+dt) & =  \tilde{\alpha}_{l, \mu}(t) + i \left\{\left( \sigma_l + i\frac{\kappa}{2} - \dfrac{\tilde{g}}{\tilde{N}}\right) \tilde{\alpha}_{l, \mu}(t) \right.\\
			&  +\tilde{g}\sum_{m,n,p}\delta_{n+l,m+p} \tilde{\alpha}_{m, \mu}(t) \bar{\tilde{\alpha}}_{n, \mu}(t) \tilde{\alpha}_{p, \mu}(t) \\
			&  \left. -\frac{\kappa}{2} \tilde{F} \delta_{l,0} \right\} dt + 
			\sqrt{\kappa \,dt/(2\tilde{N})} \chi_{l, \mu}(t) \ ,
		\end{split}
		\label{eq:TWA_Rescaled}
	\end{equation}
which explicitly depends on $\tilde{N}$.

For coherent states, the scaling parameter $\tilde{N}$ is proportional to the ratio between the field intensity and the fluctuations of the field quadratures, and therefore can be interpreted as a measure of the \textit{classicality} of the optical system. 
Small values of $\tilde{N}$ describe a regime with sizeable quantum effects, where quantum fluctuations are of the same order as the field intensity. As $\tilde{N}$ increases, fluctuations become smaller compared to the field intensity, and quantum effects become less relevant. 
This interpretation of the quantity $\tilde{N}$ holds also in the TWA, which describes quantum states beyond the coherent-state approximation. Indeed, for large values of $\tilde{N}$, Eq. \eqref{eq:TWA_Rescaled} approaches the GP equation \eqref{eq: GPE}. 
Our goal is to investigate how the quantum effects influence the dynamics of a DKS by comparing the solution of Eq. \eqref{eq:TWA_Rescaled} obtained for different values of $\tilde{N}$, while keeping all the other parameters unchanged. Notice that, in light of the scaling relations in Eq. \eqref{eq: rescaling relation}, this procedure corresponds to solving the Lindblad master equation Eq. \eqref{eq:master equation} for different values of the nonlinearity $g$ and the pump amplitude $F$, in such a way that the product $F^2 g$ remains constant.

\section{Results}
\label{sec:results}

\subsection{Regime of parameters} 

\begin{figure}
	\centering
	\includegraphics[width=1\linewidth]{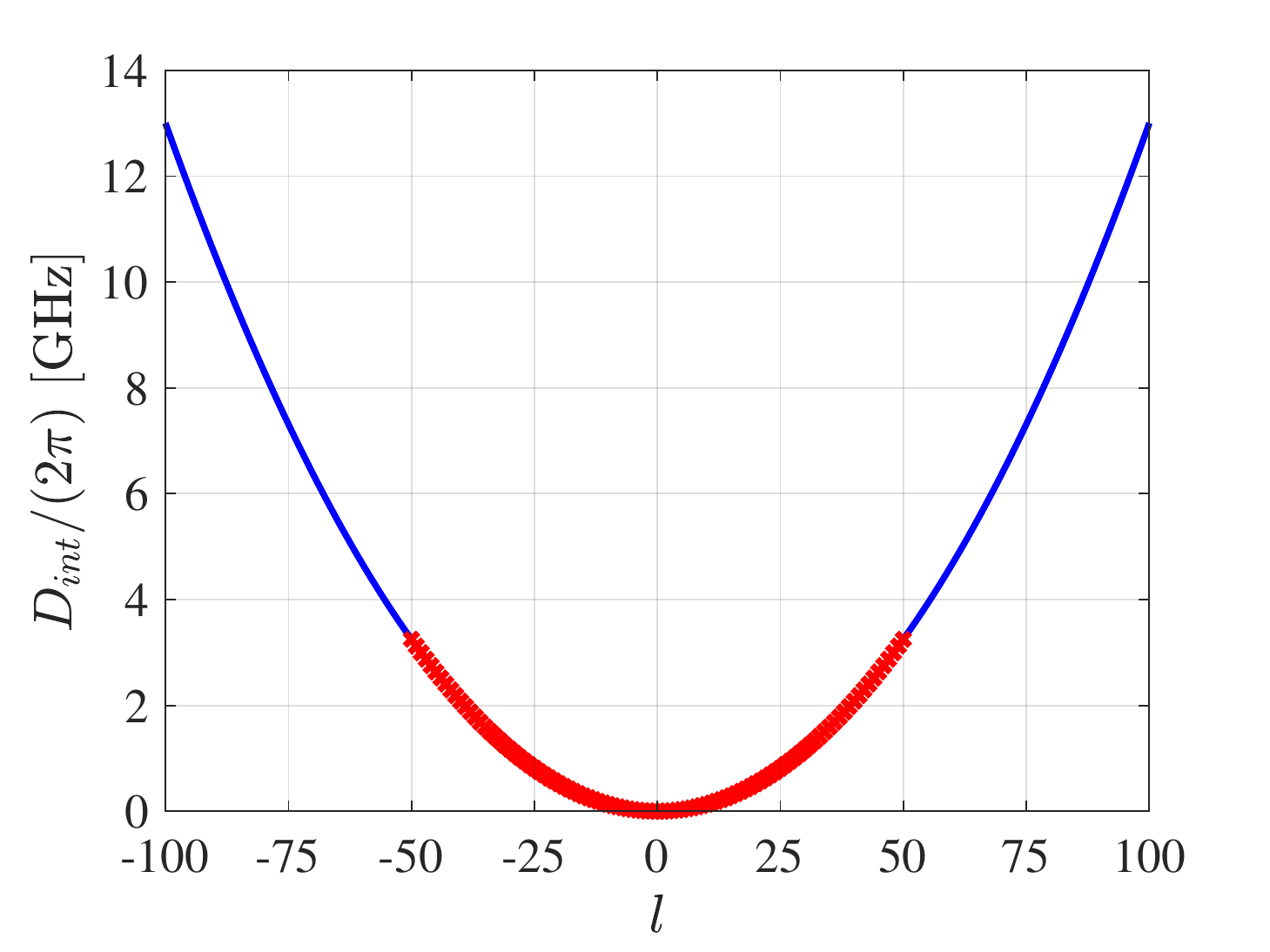}
	\caption{Integrated dispersion relation $D_{int}(l)$ versus relative mode number $l$. The red markers show the 101 spectral modes that are considered in the numerical simulations. The parameters are quoted in the text and correspond to typical experimental situations \cite{Brasch2016}.}
	\label{fig:integrated disp relation}
\end{figure}

A micro-ring resonator is characterized by its radius $R$ (and length $L=2 \pi R$), cross-section $A_{\rm eff}$, quality factor $Q$, resonant central frequency $f_0=\omega_0/2\pi$, refractive index $n_0$, the Kerr parameter $n_2$, and the group velocity dispersion $\beta_2$. The system is driven by a laser with frequency $\omega_p$.

From these quantities, the loss rate $\kappa$ and the nonlinearity $g$ in the Lindblad master equation \eqref{eq:master equation} can be determined respectively as $\kappa = \omega_0/Q$, $g = \hbar\omega_0^2c n_2/(n_0^2A_{\rm eff}L)$, where $c$ is the speed of light. Close to the driving field frequency (i.e., $\omega_0 \simeq \omega_p$), the mode dispersion of the microresonator is approximated using a second-order polynomial
\begin{equation}
	\omega_l = \omega_0 + D_1 l+ \dfrac{1}{2} D_2 l^2\,,
\end{equation}
where $D_1=c/(n_0 R)$ is the mean free spectral range (FSR) and $D_2= -(c/n_0) D_1^2 \beta_2$. A positive value of $D_2$ characterizes the anomalous dispersion regime, which is needed for the formation of DKSs \cite{Brasch2016}. The integrated dispersion relation $D_{\rm int}(l)$ relative to the driving mode at $l = 0$ is defined by  [c.f. Fig.~\ref{fig:integrated disp relation}]
\begin{equation}
	D_{\rm int}(l)\equiv \omega_l - (\omega_0+D_1 l)\;.
\end{equation}
The driving parameter $F$ is related to the power of the external driving field through the relation $P_{ext} = \hbar\omega_p\kappa F^2/(4\eta)$, where $\eta$ is the coupling efficiency (we assume critical coupling, i.e., $\eta = 1/2$).	
	
Typical parameters for a silicon nitride ($\textrm{Si}_3 \textrm{N}_4$) ring resonator encapsulated in silica \cite{Ji:17,Brasch2016}, with $R = 100~\mu$m, are $A_{\rm eff}=0.73 \cdot 2.5\cdot 10^{-12}~\text{m}^2$ \cite{Ji:17}, $Q = 1.5 \cdot 10^6$ \cite{Brasch2016}, $f_0 = \omega_0/2\pi=193.5$ THz corresponding to a wavelength $\lambda = 1.55 \, \mu$m in the telecom range, $n_0=1.99$, $n_2=2.4\cdot10^{-19}\text{ m}^2/\textrm{W}$. Consequently, $\kappa =8.1\cdot 10^{8}~\text{s}^{-1}$ and $g =2.47~\text{s}^{-1} = 3.05 \times 10^{-9} \kappa$ \cite{PainterOE2005, PhysRevApplied.12.064065, PhysRevA.93.033820}. Accordingly, the laser frequency to fulfill our condition is $\omega_p/2\pi = 193.47$ THz, i.e., a detuning $\sigma_0/2\pi = -0.132$ GHz. 

A necessary condition to observe a DKS in the solution of the GP equation is that the driving field $F$ be larger than a minimum threshold value $F_{\rm thr}$ \cite{PhysRevA.89.063814,PhysRevA.93.033820,Herr2014}. Expressing the minimum threshold condition in terms of the parameters of the present model results in $F_{\rm thr}^2g/\kappa = \tilde{F}_{\rm thr}^2\tilde{g}/\kappa = 1/2$. For the present analysis, we set $F = 1.8 \times 10^4$. This value, and the value chosen for $g$, correspond to the typical regime of current experiments, where quantum fluctuations are very small relative to the classical field. They also correspond to a regime well above the the minimal threshold for soliton formation in the classical description in terms of the GP equation~\cite{PhysRevA.89.063814}. We arbitrarily set $\tilde{N}=1$ for this choice of $F$ and $g$. In what follows, we will study the results of the TWA for values of $\tilde{N}$ ranging between $\tilde{N}= 6.3\times10^{-6}$ and $\tilde{N}=1$. Values $\tilde{N}<1$ describe cases with larger nonlinearity $\tilde{g}$ and smaller driving field amplitude $\tilde{F}$, than the values of $F$ and $g$ quoted above, for which quantum effects are larger. Finally, we define the dimensionless time $\tau=\kappa t/2$.

\begin{figure}
	\centering
	\includegraphics[width=1\linewidth]{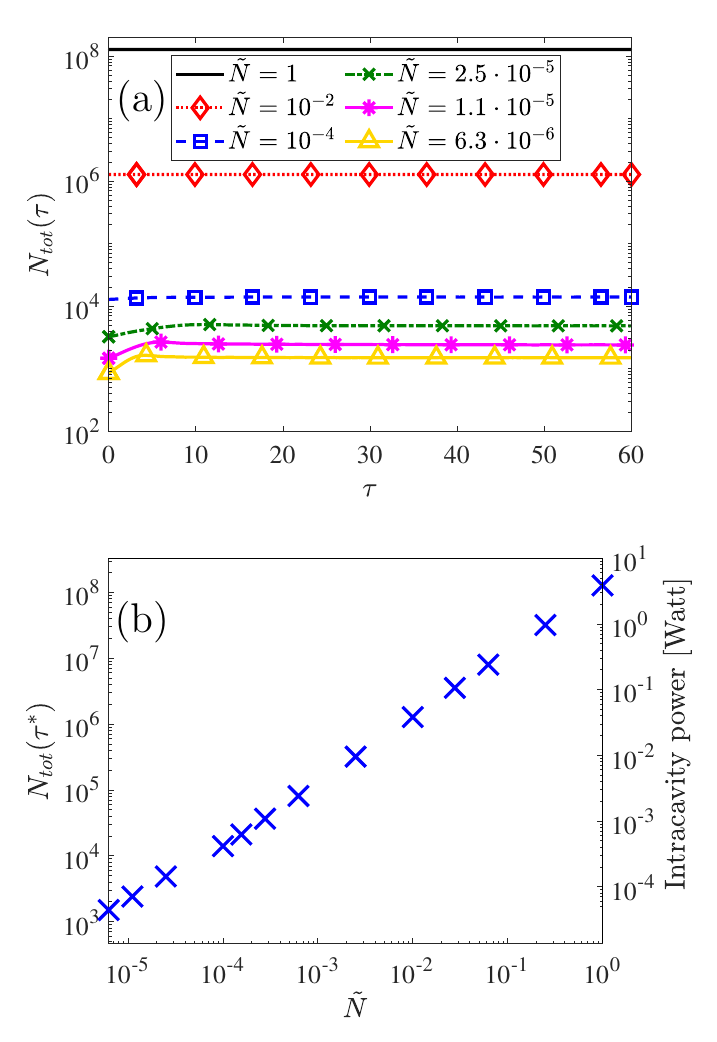}
	\caption{(a) Time evolution of the total number of photons in the ring for different values of $\tilde{N}$. (b)  Total number of photons in the ring and intracavity power versus the scaling parameter $\tilde{N}$. The values are taken at $\tau^*=60$, where the photon number reached a stationary distribution.}
	\label{fig:total pop and intracavity power}
\end{figure}

\subsection{Dynamics of the soliton}

We study the time evolution of DKSs by numerically solving Eq.~\eqref{eq:TWA_Rescaled} for the rescaled fields $\tilde{\alpha}_l$ obtained using the TWA approach. Eq.~\eqref{eq:TWA_Rescaled} gives rise to a stochastic trajectory in the space of the fields $\tilde{\alpha}_{l}$, determined by the specific realization of the noise term $\chi_{l,\mu}(t)$. All results in what follows are obtained by averaging over several trajectories arising from different realizations of $\chi_{l,\mu}(t)$.
As initial conditions, we assume each mode to be in a coherent state corresponding to the solution $\tilde{\alpha}_{l}^{\rm GPE}$ of the GP equation, which in turn is obtained by numerically integrating Eq.~\eqref{eq: GPE} at long times. This choice has the advantage of avoiding the integration of a possibly long transient before the actual formation of a soliton.
In the TWA formalism, this choice of initial condition implies that the initial condition $\tilde{\alpha}_l(t=0)$ in Eq.~\eqref{eq:TWA_Rescaled} must be sampled from a Gaussian distribution of variance $1/(2\tilde{N})$ and average $\tilde{\alpha}_{l}^{\rm GPE}$.

More precisely, we set
\begin{equation}
		\tilde{\alpha}_{l, \mu}(t=0) = \tilde{\alpha}_{l}^{\rm GPE}+\frac{1}{\sqrt{2 \tilde{N}}}\eta_{l, \mu}\;,
		\label{eq: init condition TW}
\end{equation}
where $\eta_{l, \mu}$ is a complex random variable of zero mean, verifying $\mean{\eta_{l, \mu}\eta_{l, \mu'}}=0$ and $\mean{\bar{\eta}_{l, \mu}\eta_{l, \mu'}}=\delta_{\mu,\mu'}$.

Fig.~\ref{fig:total pop and intracavity power}(a) displays the total number of photons in the ring $N_{\rm tot}$ vs time, for varying $\tilde{N}$. The small initial transient is due to the difference between the solution of the GP equation, which was used as initial condition, and the actual TWA solution.
In what follows, when analyzing spectral features of the DKS, data will be taken in the vicinity of $\tau^*=60$, which are not affected by the transient.
In Fig. \ref{fig:total pop and intracavity power}(b) the dependence of $N_{tot}(\tau^*)$ on $\tilde{N}$ is shown to be linear, $N_{tot} \sim \tilde{N}$. Thus, also the intracavity power $P_I = \hbar\omega_p D_1 N_{tot}/(2 \pi)$ (which is proportional to the total number of photons in the microring) depends linearly on $\tilde{N}$. We conclude that a small photon occupation and low intracavity power is reached only for small $\tilde{N}$.

\begin{figure}
	\centering
	\includegraphics[width=1\linewidth]{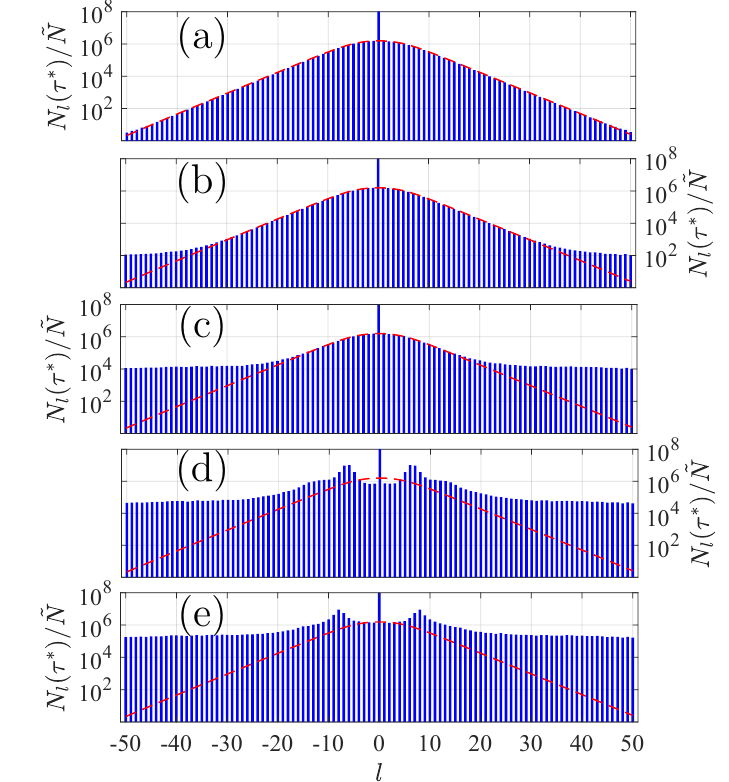}
	
	\caption{Snapshot of the mode occupation at $\tau^*$ for $\tilde{N} = 1$ (a), $10^{-2}$ (b), $10^{-4}$ (c), $2.5\cdot 10^{-5}$ (d), and $6.3\cdot 10^{-6}$ (e). The red dashed lines show the GPE prediction for the mode occupation. Parameter values: $\sigma_0=-1.024\kappa$, $D_1=1.8587\cdot 10^3 \kappa$, $D_2 = 2.02\cdot 10^{-2}\kappa$, $\tilde{g} = 3.05\cdot 10^{-9}\kappa$ and $\tilde{F} = 1.8\cdot 10^4$.}
	\label{fig:soliton k space}
\end{figure}

In Fig.~\ref{fig:soliton k space}, the photon number in the $l$-th mode $N_l(\tau^*)$ [c.f. Eq.~\eqref{eq:TWAPhotonsMode_l}], is displayed for different values of $\tilde{N}$. For the largest value of $\tilde{N}$, the output field is in agreement with the prediction of the GP equation, while smaller values of $\tilde{N}$ gradually display increasing features of quantum fluctuations.

\begin{figure*}
	\centering%
	\includegraphics[width=1\textwidth]{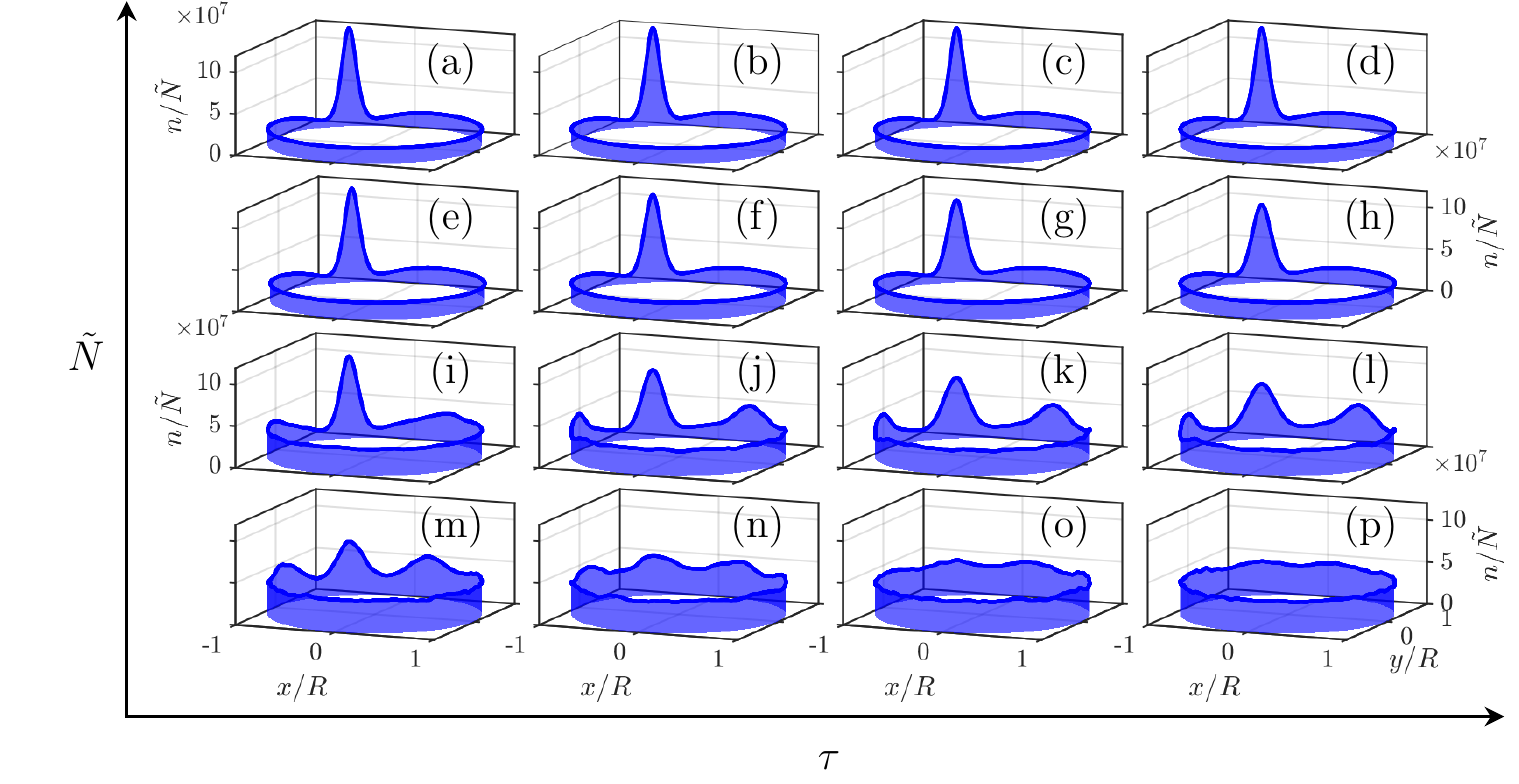}
	
	\caption{Snapshot of the field density in real space at $\tau=0.5\cdot10^4\kappa/2 T\approx 8.5$ (a,e,i,m), $\tau=1.5\cdot10^4\kappa/2 T\approx 25.4$ (b,f,j,n), $\tau=2.5\cdot10^4\kappa/2 T\approx 42.3$ (c,g,k,o), and $\tau=3.5\cdot10^4\kappa/2 T\approx 59.2$ (d,h,l,p), for $\tilde{N} = 10^{-2}$ (a,b,c,d), $ 10^{-4}$ (e,f,g,h), $2.5\cdot 10^{-5}$ (i,j,k,l), and $6.3\cdot 10^{-6}$ (m,n,o,p). The density is plotted as a function of the coordinates $x = R \cos(\theta)$ and $y = R \sin(\theta)$, $R$ being the radius of the ring resonator. The soliton is depicted at times multiple of $T=2\pi/D_1\approx 4.2\cdot 10^{-12}$ s, the rotation period of the soliton along the ring. For this choice of time, the peak of the soliton always occupies the same position in ring, allowing an easier comparison between the different plots.}
	\label{fig: field density time snapshots 3D}
\end{figure*}

In Fig.~\ref{fig: field density time snapshots 3D} the photon density along the ring $n(\theta,\tau) = N_\theta(\tau)/(2\pi)$, $N_\theta$ being the number of photons in the position $\theta$ [c.f. Eq.~\eqref{eq:TWAPhotonsrealspace}]\footnote{Notice that this definition ensures that $N_{\rm tot}(\tau)=\sum_l N_l (\tau) = \int\text{d} \theta \, n(\theta, \tau)$.} , is displayed at increasing times $\tau$ (left-to-right) and increasing $\tilde{N}$ (bottom-to-top). 
For $\tilde{N}=1$ in Fig.~\ref{fig: field density time snapshots 3D}(a-d), the soliton displays a constant profile within the considered time window.
However, for smaller $\tilde{N}$, i.e., increasing the relevance of quantum fluctuations, the soliton profile changes in time, in particular by showing a decreasing contrast of the intensity profile along the ring.
For the smallest value $\tilde{N} = 6.3\cdot 10^{-6}$, Fig.~\ref{fig: field density time snapshots 3D}(m-p), the photon density quickly approaches a uniform distribution over the ring. We conclude that the soliton is gradually smeared out over time by quantum fluctuations, and smaller values of $\tilde{N}$ correspond to faster disappearance of the soliton.

    To quantify the soliton lifetime, we compute the Liouvillian gap $\Lambda$ (i.e., the slowest decay rate).
	Indeed, the DKS is the longest-lived process of Eq.~\eqref{eq:master equation}, and thus $\Lambda$ is the inverse of soliton lifetime (see Sec.~\ref{subsec:ME and Liouvillian gap}).
	To extract $\Lambda$, we consider the time evolution of the \textit{contrast} of the soliton defined by 
	\begin{equation}
		C(\tau)=\dfrac{\text{max}_\theta(n(\theta,\tau))}{\int_0^{2\pi}n(\theta,\tau)d\theta / 2\pi}\,.
		\label{eq:def contrast}
	\end{equation}
For a flat intensity profile along the ring, the value of $C(\tau)$ approaches 1. We estimate $\Lambda$ by assuming an exponential behaviour vs time, $C(\tau) \simeq  1 + A \exp(-\Lambda \tau)$ and fitting the numerical results. 

\begin{figure}
	\centering
	\includegraphics[width=1\linewidth]{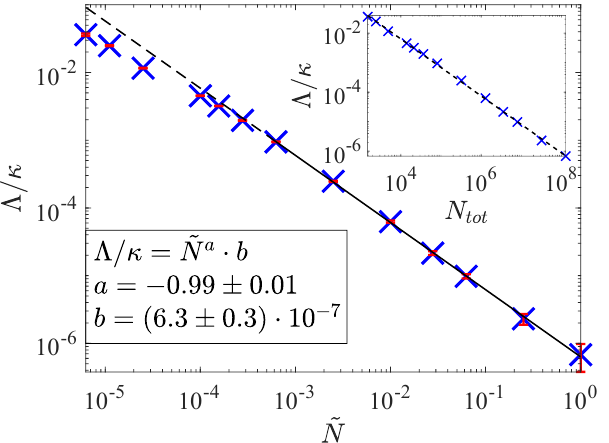}
	\caption{(a) Liouvillian gap $\Lambda$ versus $\tilde{N}$. The error bars of the fit (in red) show the standard error (the 95\% confidence interval) of each point. The power-law fit of the Liouvillian gap for $\tilde{N}\geq 6.3\cdot 10^{-4}$ is shown by the continuous line. The coefficients of the fit are $\Lambda/\kappa=\tilde{N}^a\cdot b$ with $a=-0.99\pm0.01$ and $b=(6.3\pm0.3)\cdot 10^{-7}$. Inset: the Liouvillian gap versus the total number of photon inside the ring microresonator at long times (see Fig.~\ref{fig:total pop and intracavity power}). The dashed line represents a power law fit.}
	\label{fig:Liouv gap vs N_tilde}
\end{figure}

In Fig.~\ref{fig:Liouv gap vs N_tilde}, the Liouvillian gap is plotted as a function of $\tilde{N}$. For large $\tilde{N}$ the Liouvillian gap follows a power law $\Lambda \sim \tilde{N}^a$ with $a<0$, indicating that the gap closes in the classical limit $\tilde{N} \to \infty$. A similar power law emerges in the dependence of $\Lambda$ on $N_{\rm tot}$ (inset of Fig. Fig.~\ref{fig:Liouv gap vs N_tilde}). Here, $\Lambda \sim N_{\rm tot}^\eta$ with $\eta = -0.97 \pm 0.01$. This analysis indicates the range of values of the input power for which a finite soliton lifetime may be observed.

\subsection{DKS as a dissipative time crystal}
\label{sec:scheme spectrum DKS}

\begin{figure}[ht]
	\centering
	\includegraphics[width=1\linewidth]{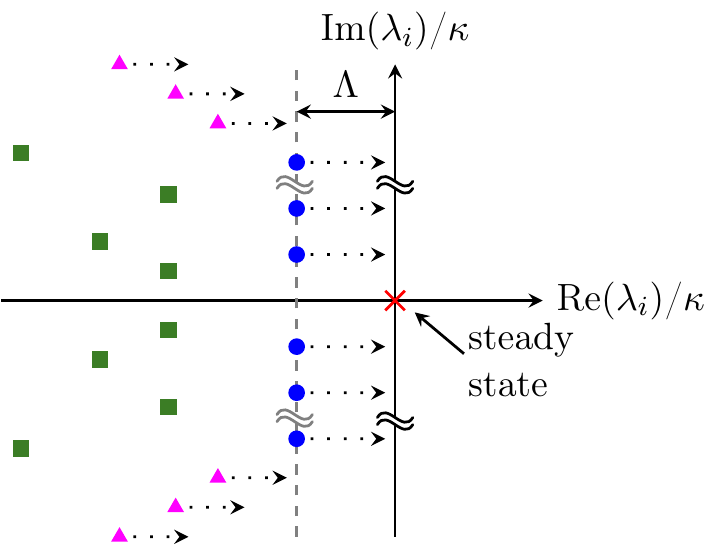}
	\caption{Schematic representation of the spectrum of the Liouvillian. The spectrum always has one zero eigenvalue, corresponding to the steady state (red cross). A set of eigenvalues with vanishing real part and equally-spaced imaginary parts emerges in the classical limit of large $\tilde{N}$ (blue circles). The Liouvillian gap is the distance between the complex eigenvalues with the largest real part and the imaginary axis.}
	\label{fig:scheme spectrum DKS}
\end{figure}

The occurrence of a time-crystalline phase in a dissipative system is signaled by the emergence of several eigenvalues of the Liouvillian whose real part tends to zero in the thermodynamic limit, and whose imaginary part is a multiple of a finite frequency, as schematically shown in Fig.~\ref{fig:scheme spectrum DKS}.

We extract the imaginary part of the Liouvillian eigenvalues with the largest real part, by studying the Fourier spectrum of the KFC
\begin{equation}
	S_\varphi(\omega) = \left|\dfrac{\sqrt{2\pi}}{N_{\Tau} T}\int_{t_0}^{t_0+N_{\Tau} T}dt \me^{i\omega t} \varphi (\theta =0, t)\cdot\tilde{N}\right|^2\;.
	\label{eq:power spectrum WKT}
\end{equation}
where $\varphi(\theta,t) = \textrm{Tr} \left[ \op{\rho}(t) \times 1/\sqrt{2\pi}\sum_l e^{i\theta l}\ao_l\right]$, and $T=2\pi/D_1$ is the rotation period of the soliton along the ring.
The parameters $t_0 = 20\kappa^{-1}$ and $N_{\Tau} = 2\cdot 10^4$ are set to ensure that the dynamics is dominated by the eigenvalues with the largest real part. 
Notice that, for these parameters, $S_\varphi(\omega)$ does not depend significantly on the position $\theta$ at which the field $\varphi$ is considered. 

\begin{figure}
	\centering
	\includegraphics[width=1\linewidth]{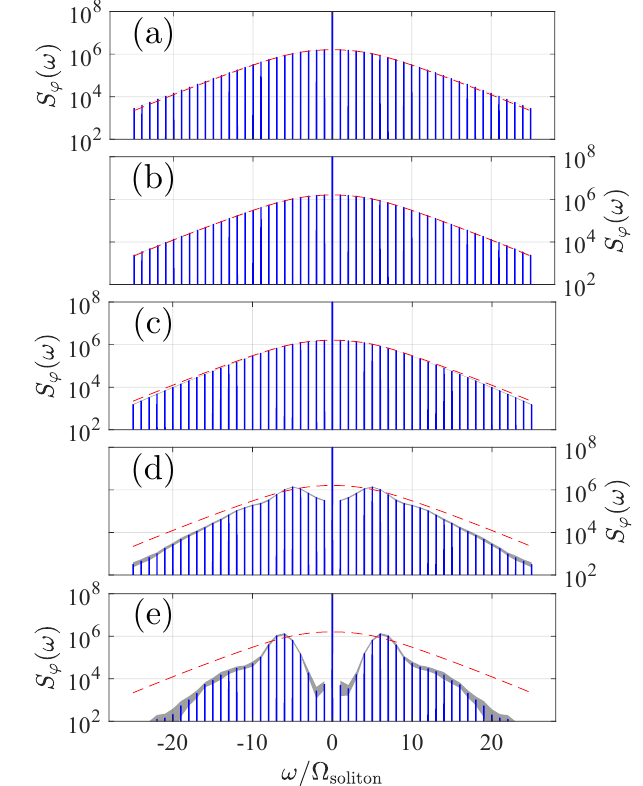}
	\caption{Fourier spectrum obtained by considering 51 modes and for $\tilde{N} = 1$ (a), $\tilde{N} = 4\cdot 10^{-4}$ (b), $\tilde{N} = 10^{-4}$ (c), $\tilde{N} = 2.5\cdot 10^{-5}$ (d) and $\tilde{N} = 1.1\cdot 10^{-5}$ (e). The gray region shows the $95\%$ standard deviation. The red dashed lines represent the envelope of the Fourier spectrum obtained from the GPE.
	}
	\label{fig:power spectra}
\end{figure}

The computed power spectra are plotted in Fig.~\ref{fig:power spectra}. 
From the spectra, we extract the frequency spacing $\Omega_{\text{soliton}} = D_1 = 1.86 \cdot 10^3 \kappa$, which coincides with the classical prediction, indicating that quantum fluctuations affect mainly the lifetime of the soliton, while having negligible effects on the period of its motion along the ring. For the smallest value of $\tilde{N} = 6.3\cdot 10^{-6}$, where $\Lambda \simeq 3 \times 10^{-2} \kappa$, the ratio $\Lambda/\Omega_{\text{soliton}}$ is of the order of $10^{-4}$. These results indicate that the peculiar structure of the frequency comb (i.e. the presence of equally spaced, narrow spectral lines) is preserved even in the regime where quantum effects produce a significant departure from the predictions of the classical GP equation.

\section{Conclusions}
\label{sec:conclusions}

We have carried out a theoretical study of DKSs in microring resonators in terms of the truncated Wigner approximation, which describes quantum fluctuations to leading order in $\hbar$ and is therefore well suited for the description of regimes of large photon occupation as in current experiments. We have shown that quantum effects are responsible of a finite lifetime of the soliton, which in the long time limit leaves place to a solution with the field uniformly distributed along the ring. The timescale of the soliton decay depends on the relative size of quantum fluctuations, and decreases when quantum fluctuations become larger. A scaling analysis of the TWA equations indicates that a regime with large quantum effects may be achieved by decreasing the driving field intensity while correspondingly increasing the strength of the Kerr nonlinearity. The analysis provides clear indications about whether this behaviour can be observed in experiments.

We have additionally shown that the timescale associated with the soliton disappearance is determined by the inverse Liouvillian spectral gap. More precisely, by studying the power spectrum of the DKS, we have inferred the complex eigenvalues of the Liouvillian super-operator which governs the dynamics of the DKS as an open quantum system. We have shown that the eigenvalues with the largest real part -- besides the zero-eigenvalue associated to the spatially uniform steady state -- are arranged to have a constant (negative) real part, defining the Liouvillian gap, and evenly spaced imaginary parts, corresponding to the Kerr frequency comb. This arrangement emerges asymptotically in the limit of large input power, and the Liouvillian gap vanishes as a power law of the total photon occupation in the microring modes. We have therefore shown that DKSs are a specific manifestation of a dissipative time crystal -- a general phenomenon which can arise in open quantum systems and has been extensively studied in recent times. Establishing the link between DKSs and dissipative time crystals is an important step in the study and characterization of spontaneous time-translational symmetry breaking in quantum systems out of equilibrium.

\section*{Acknowledgments}
We acknowledge enlightening discussions with Juan Pablo Vasco. This work was supported by the Swiss National Science Foundation through Project No. 200021\_162357 and 200020\_185015.
	
\clearpage	
	
\bibliographystyle{FabrizioStyle}
\bibliography{biblatex-soliton}

\end{document}